\documentclass[12pt,a4paper]{article}
\usepackage{mathtext}
\usepackage[cp1251]{inputenc}
\usepackage[T2A]{fontenc}
\usepackage {graphicx}
\usepackage[russian,english]{babel}
\newcommand{\bm}{\mathbf}

\title{Feasibility of finite renormalization of particle mass in quantum
electrodynamics}
\author{A.V. Gichuk,V.P. Neznamov, Yu.V. Petrov\\
Russian Federal Nuclear Center, Sarov, Russia, 607190}
\date{\today}
\begin{document}
\maketitle

\begin{abstract}
The paper proposes an algorithm for regularization of the self-energy
expressions for a Dirac particle that meets the relativistic and gauge
invariance requirements.

Within the second order formulations of the ``old'' perturbation theory for free
motion, the expression for the upper integration limit $q$ is a slowly
enough varying function of particle impulse. For a particle at
rest, $q=m$; for $p/m$ in the range from 0 to 1,
\[
q \approx m\left( {1 + 0.00864\frac{{p^{4}}}{{m^{4}}} -
0.00217\frac{{p^{6}}}{{m^{6}}}} \right);
\]
for the ultra-relativistic case $|\bm{p}| \gg m:\quad q \approx 1.5m$.

For 4D perturbation theory, on introduction of the limiting 4-impulse,
$L^2 = L_0^2 - \bm{L}^2$, it is shown that with a large time
component, $L_0/m\gg 1$, the spatial values of
$L_i$ are limited and are the same as the components of the
introduced limits of integration $q_i$:
$\bm{L}^2 = \bm{q}^2$.

Within the proposed algorithm, in the second-order of the perturbation
theory, the renormalized Dirac particle mass is
\[
m^{\left( {2} \right)} = m_{0} + \Delta m^{\left( {2} \right)} = m_{\,0}
\,\left( {1 + 1.115\frac{{e^{2}}}{{\pi} }} \right),
\]
where $m_0$ is the bare mass of the particle.

The electromagnetic radius of particle at rest determined by equating the
self- energy and the electrostatic energy $e^2/r_{em}$ is about half the
Compton wavelength, $r_{em} \approx 1/2m$.
\end{abstract}
\newpage

It is commonly known that the calculation of some physical effects results
in undefined diverging expressions, when the perturbation theory formalism
of existing formulations of field and particle quantum theories is used. If
the quantum-field theories are renormalizable, the undefined expressions are
removed in all orders of the perturbation theory by renormalizing mass and
charge of relevant particles. In particular, quantum electrodynamics is such
a theory.

In the lifetime of the relativistic quantum-field theories, many researchers
tried to solve the regularization problem of infinite expressions appearing
in the computations from various viewpoints. The problem, however, remains
unsolved.

This paper suggests, as its authors believe, a natural way for solving the
problem of finite renormalization of particle mass by the example of quantum
elecrodynamics.

First, use the formulations from the ``old'' perturbation theory given by
Dirac for quantum electrodynamics in \cite{ref1}.

The self-energy operator in the second order of the perturbation theory is:

\begin{equation}
\label{eq1}
Y = \frac{e^2}\pi \frac 1{4\pi}
\int\alpha_\mu
\frac{\nu + \tilde{E}/|\tilde{E}|}{|\tilde{E}|+|{\bm k}|-\nu E_i}
\alpha^\mu \frac{d{\bm k}}{|{\bm k}|}
\end{equation}

In (\ref{eq1}) and hereinafter
$\hbar=c=1;\quad \alpha^\mu\left\{
\begin{array}{l}
1 \\
\alpha^i\\
\end{array}\right.$
\\
$\alpha^i,\, \beta$ are Dirac matrices; $\bm{\tilde{p}} = \bm{p} + \bm{k};$

\begin{eqnarray*}
&&E=\bm{\alpha}\bm{\pi} + \beta m + eA_0;\quad \pi^i = p^i - eA^i;\quad
\tilde{\pi}^i = p^i + k^i - eA^i;\\
&& \tilde {E} = \bm{\alpha} \tilde{\bm \pi} + \beta m + eA_0;  \quad
p_0 = \left(m^2 + {\bm p}^2\right)^{1/2};\quad
\tilde{p}_0 = \left(m^2 + \bm{p}^2 + 2\bm{p}\bm{k} +
{\bm k}^2 \right)^{1/2};
\end{eqnarray*}
$A_0,\, A^i$ are scalar and vector potentials of the external
electromagnetic field.

Expression (\ref{eq1}) implies the averaging over
$\nu : (1/2)\left(Y(\nu = 1) + Y(\nu = - 1)\right)$.

In action on state $\left|i\right\rangle$ with energy $E_i$,

\begin{equation}
\label{eq2}
E\left| i \right\rangle = E_{i} \left| i \right\rangle\,, \quad
Y\left| i \right\rangle = Y_{i} \left| i \right\rangle
\end{equation}

Dirac considered three modes for the regularization of expression (\ref{eq1})
in \cite{ref1}.

\underline{Mode {\it a}}.
Modulus of $\bm{k}$ is limited, $|{\bm k}|<q$, where $q$ is some number.

\underline{Mode {\it b}}.
Here the sum of energy moduli of three particles involved in the interaction event
is limited, $|\tilde{E}| + |{\bm k}| + |E_i| < 2q$.

\underline {Mode {\it c}}.
Here the sum of kinetic energies of three particles is limited,
\[
|\tilde{E} - eA_0| + |{\bm k}| + |E_i- eA_0| < 2q.
\]

For large $q$, the correspondence is established:

\begin{eqnarray}
\label{eq3}
&&Y_a = Y_{b} + \frac 16 \bm{\alpha}\bm{\pi} + \frac 12 eA_0; \nonumber\\
&&Y_a = Y_{c} + \frac 16 \bm{\alpha}\bm{\pi}
\end{eqnarray}

As a result, Dirac prefers \underline{mode {\it c}} as a technique
allowing us to eventually obtain relativistically- and
gauge-invariant expressions for the self-energy and vacuum polarization. As
this is always the case in the relativistic quantum field theories, the
numerical value of $q$ is not determined.

This paper uses \underline {mode {\it a}} to
demonstrate the possibility to obtain the relativistically- and
gauge-invariant expression for the self-energy with simultaneous estimation
of the upper integration limit $q$.

First, consider the case where there are no external electromagnetic fields
$(A_0 = 0,\,A^i = 0)$. Expression (\ref{eq1}) becomes

\begin{equation}
\label{eq4}
Y_0 = \frac{e^2}{4\pi^2}\int \alpha _\mu
\frac{\nu\tilde{p}_0 + {\bm \alpha} \bm{\tilde{p}} + \beta  m}
{\tilde{p}_0(\tilde {p}_0 + |{\bm k}| - \nu p_0 )}
\alpha^\mu \frac{d{\bm k}}{|{\bm k}|}.
\end{equation}

Consider a non-relativistic charged particle having impulse $|\bm{p}| \ll m$.

On summation over $\mu $ and integration over azimuthal angle $\varphi $, we
obtain

\begin{equation}
\label{eq5}
Y_0 = \frac{e^2}\pi\int\limits_0^q \int\limits_{-1}^1
\frac
{-\nu\tilde{p}_0 + {\bm \alpha}(\bm{p} + \bm{k}) + 2\beta m}
{\tilde{p}_0 (\tilde{p}_0 + |{\bm k}| - \nu p_0)}
k d(\cos\theta)dk.
\end{equation}

Expanding the integrand in (\ref{eq5}) over the particle impulse $|{\bm p}|$
up to quadratic expressions ${\bm p}^2/m^2$ and integrating give:
\begin{eqnarray*}
Y_0 & = &\frac{2e^2}\pi\left\{ m\left[
-{\ln(x)\over 4} +\frac 18 \frac{m^2}{x^2}+\frac{{\bm p}^2}{2m^2}\left(
\frac 13 \frac{m^2}{x^2 + 1} - \frac{\ln(x)}4 - \frac 1{24}\frac{m^2}{x^2}
\right)\right]\right. \\
&& +{\bm{\alpha p}}\left(\frac{\ln(x)}4-\frac{m^2}{3(x^2+1)}-
\frac 1{24}\frac{m^2}{x^2}\right) \\
&&+\left.\left.\beta m \left[ \ln(x) + \frac{\bm{p}^2}{2m^2}\left(-\frac 43
\frac{m^2}{x^2 + 1} + \frac{4m^2}{(x^2 + 1)^2} -
\frac 83\frac{m^6}{(x^2 + 1)^3}\right)\right] \right\}\right|_m^{A\cdot m},
\\
&& x=k+\sqrt{m^2+k^2},\qquad A\cdot m=q+\sqrt{m^2+q^2}.
\end{eqnarray*}

Use equalities
\[
\left( -m - \frac{{\bm p}^2}{2m} + {\bm\alpha}{\bm p} + \beta m
\right) \left| i \right\rangle \approx 0,\quad \left\langle i
\right|\bm{\alpha}\bm{p}\left| i\right\rangle = \left\langle i
\right|\frac{\bm{p}^2}{p_0} \left| i\right\rangle \approx
\left\langle i \right|\frac{\bm{p}^2}m\left| {i} \right\rangle
\]
to obtain

\begin{eqnarray}
\label{eq6}
\left\langle i \right| Y_0 \left| i \right\rangle &=&
\frac{e^2}\pi\left\langle i \right|\left\{
\beta m\left( \frac 32 \ln(A) + \frac 1{4A^2} - \frac 14\right)\right.
\nonumber\\
&&+\left.\frac{\bm{p}^2}m\left( -\frac 5{3(A^2 + 1)} +
\frac 4{(A^2 + 1)^2} - \frac 8{3(A^2 + 1)^3} + \frac 16 \right)\right\}
\left| i \right\rangle.
\end{eqnarray}

Reasoning from the relativistic invariance condition for the self-energy
operator $Y_0$, require that the coefficient of $\bm{p}^2/m$ in
(\ref{eq6}) vanish. The natural condition for the upper limit of
integration $q$ in (\ref{eq5}) follows herefrom:
\begin{eqnarray}\label{eq7}
&&-\frac 5{3(A^2 + 1)}+\frac 4{(A^2+1)^2} - \frac 8{3(A^2+1)^3} +
\frac 16 = 0 \qquad \mbox{or} \nonumber\\
&&\frac q{(q^2+m^2)^{1/2}}\left(\frac {q^2}{q^2 + m^2}-\frac 12\right)=0.
\end{eqnarray}

The solution to equation (\ref{eq7}) is $q = \pm m$, in addition to the trivial
one, $q=0$.

The value $q \approx m$ means that the upper integration limit imposes the
restriction on the distances. They should be longer than Compton wavelength
of Dirac charged particle: $|\bm{x}|\ge 1/m$, which is quite reasonable from
the viewpoint of quantum mechanics.

The value $q \approx m$ for the upper integration limit is also reasonable
in terms of ''Zitterbewegung'' effect for Dirac particle. In fact, if Dirac
Hamiltonian is written in the form
\[
H_D=\bm{\alpha}\bm{\pi} + \beta m + eA_0 ;\quad \bm{\pi} = \bm{p}-e\bm{A}
\]
in the presence of external fields, then the operator of coordinate
$x^i$ can be represented as follows \cite{ref2}:
\[
x^i = x_0^i + \pi^i\frac 1{H_D - eA_0}t -
\frac 14\alpha_0^i e^{-2i(H_D - eA_0)t}\frac 1{(H_D-eA_0)^2},
\]
where $\alpha_0^i$ is the matrix integration constant.

The third term of $x^i$ operator highly oscillates with frequency
$\sim 2m$. This term can be represented as
\begin{equation}
\label{eq8}
\Delta x^i = \frac i2\left(\alpha^i-\pi^i\frac 1{H_D - eA_0}\right)\frac 1{H_D - eA_0}
\end{equation}

It is seen from (\ref{eq8}) that $\Delta x^i\sim 1/m$, and from this
standpoint it is quite natural to restrict the consideration to distances
$|\bm{x}| \ge 1/m$.

Note that when previously estimating the self-energy of the non-relativistic
charged particle in the self-field electrodynamics in Foldy-Wouthuysen
representation, one of the authors also came to the necessity of the finite
upper integration limit with $\Lambda \approx m$ \cite{ref3}.

Clear, the value $q \approx m$ obtained from the expansion of (\ref{eq5}) to
quadratic values $\bm{p}^2/m^2$ can vary when the following
expansion terms are included. The value of $q$ can also change with external
electromagnetic fields $A^\mu(\bm{x})$ introduced.

Consider the function
\begin{equation}
\label{eq9}
\left\langle i \right|F(p,q,m)\left| i \right\rangle =
\left\langle i \right|Y_0 \left| i \right\rangle - \frac {e^2}{\pi}
\left\langle i \right|\beta m \left( \frac 32\ln(A_0)+
\frac 1{4A_0^2}-\frac 14 \right)\left| {i} \right\rangle,
\end{equation}
where $A_0 = \left.A\right|_{q = m} = 1 + \sqrt {2}$.

The second addend in (\ref{eq9}) is relativistically invariant. In its form, the
addend corresponds to the infinite renormalization term to be added to the
particle mass in standard quantum electrodynamics, if the integration limit
$q \to \infty $ in $A(q)$.

The function $F(p,m,q)$ is the relativistically non-invariant
expression that (with our approach) should vanish with an appropriate choice
of the upper integration limit $q$.

Table \ref{tab1} gives $q/m$ versus $p/m$, for which
$F(p,m,q)=0$.

\begin{table*}[h]
\caption{The upper integration limit $q(p)$} \label{tab1}
\small
\begin{tabular*}{\textwidth}{c|*{6}{@{\extracolsep{\fill}}c}}
\hline
 p/m & 0& 0.01& 0.02& 0.04& 0.05& 0.06\\ q/m  &1
&1.000000000088 &1.0000000014   & 1.000000023 & 1.000000055&
1.00000011\\
 \hline
 p/m  &0.07& 0.08 &0.09 &0.1  &0.2   &0.5 \\
 q/m   &1.00000021    & 1.00000036
  &1.00000058& 1.00000088   &1.000014  & 1.000512\\
  \hline
 p/m & 0.8& 1& 10& 100& 500& 1000 \\
  q/m &1.00297  & 1.00647   & 1.26716   & 1.448 & 1.48117   &
  1.494\\
  \hline
\end{tabular*}
\normalsize
\end{table*}

The following equalities were used in the numerical computations:
\[
\left\langle i \right|\beta m \left| i \right\rangle =
\left\langle i \right|\left( \frac{m^2}{(m^2 + \bm{p}^2)^{1/2}}\right)
\left| i \right\rangle,\qquad
\left\langle i \right| \bm{\alpha}\bm{p}\left| i \right\rangle =
\left\langle i \right|\left( \frac{\bm{p}^2}{(m^2 + \bm{p}^2)^{1/2}}\right)
\left| i \right\rangle.
\]

The computed data shows that the upper integration limit $q$ is a
slowly enough varying function of $p/m$. Within the range of
$p/m$ values from 0 to 1, the function is amenable to
approximation by the polynomial expression
\begin{equation}
\label{eq10}
q \approx m\left(1 + 0.00864\frac{p^4}{m^4} -
0.00217\frac{p^6}{m^6}\right).
\end{equation}

For the ultra-relativistic case of $|\bm{p}|\gg m$, the function approaches
$q \approx 1.5m$ asymptote.

Thus, when choosing the upper integration limit $q(p)$ according
to Table \ref{tab1} for free motion, the average
value of the self-energy operator (\ref{eq4}) is a relativistic invariant,
\begin{equation}
\label{eq11}
\left\langle i\right|Y_0 \left| i \right\rangle =
\frac {e^2}\pi
\left\langle i \right| \beta m \left( \frac 32\ln(A_0) +
\frac 1{4A_0^2} - \frac 14\right)
\left| i \right\rangle = 1.115\frac{e^2}\pi
\left\langle i \right|\beta m\left| i \right\rangle.
\end{equation}

We now turn to the modern 4D relativistically invariant formulation of the
perturbation theory in the quantum field theory.

The self-energy operator (or mass operator) in the second-order of the
perturbation theory is
\begin{equation}
\label{eq12}
M(p) = - \frac{8\pi i}{(2\pi)^4}e^2
\int \frac{2m - \hat{p} + \hat{k}}{[(p-k)^2 - m^2] k^2}d^4k.
\end{equation}

For the external electron line of Feynman diagram,
\[
p^2 = p_0^2 -\bm{p}^2 = m^2,\quad \quad \hat{p}\left| i \right\rangle =
m\left| i \right\rangle.
\]

Introduce, according to \cite{ref4}, the notion of the limiting 4-impulse,
$L^2 = L_0^2 - \bm{L}^2$, and perform the integration in (\ref{eq12})
over the variable $k^0$ using the Feynman rule of pole bypass.
For $L_0/m\gg 1$ the integration result is
\begin{eqnarray*}
M(p)& =& - \frac{e^2}{4\pi^2}\beta
\int d\bm{k}\left[
-\frac{|\bm{k}| - \bm{\alpha}\bm{k} + \beta m}
{|\bm{k}|(\tilde{p}_0 + |\bm{k}| - p_0)
(\tilde{p}_0 - |\bm{k}| + p_0)} \right.\\
&&+\frac{|\bm{k}| + \bm{\alpha}\bm{k} - \beta m}
{|\bm{k}|(\tilde{p}_0 - |\bm{k}| - p_0)
(\tilde{p}_0 + |\bm{k}| + p_0)}+
\frac{(p+\tilde{p}_0) - \bm{\alpha}\bm{k} + \beta m}
{\tilde{p}_0(\tilde{p}_0 + |\bm{k}| + p_0)
(\tilde{p}_0 - |\bm{k}| + p_0)}\\
&&+\left.\frac{(p-\tilde{p}_0) - \bm{\alpha}\bm{k} + \beta m}
{\tilde{p}_0(\tilde{p}_0 + |\bm{k}| - p_0)
(\tilde{p}_0 - |\bm{k}| - p_0)}\right].
\end{eqnarray*}
It can be shown that on the algebraic transformations the average of the
expression in state $\left| i \right\rangle $ equals to that of expression
(\ref{eq4}) for $Y_0$:
\[
\left\langle i \right|M(p)\left| i \right\rangle =
\left\langle i \right|Y_0 \left| i \right\rangle .
\]
Hence, the spatial components of the introduced limiting impulse $L$
are equal, $L_i = q_i$. Since for free motion the only relativistic
invariant is particle mass $m$, then $L^2 = L_0^2 - \bm{q}^2= C \cdot m^2$,
where $C$ is a numerical factor. As $L_0/m \gg 1$, the factor $C$
should be also much higher than one. With $q$ varying as a function of the
particle impulse, the time component $L_0$ should also vary accordingly, with
the invariant $L^2$ remaining invariable.

Note that if the 4D integration in (\ref{eq12}) is performed in the conventional
manner, then expression (\ref{eq12}) can be written as
\begin{equation}
\label{eq13}
M(p) = \frac{e^2}{\pi}m\left[\frac 34\ln\left(1 + \frac{L^{\prime\,2}}
{m^2}\right) + \frac{L^\prime}{2m}\arctan\frac m{L^\prime} -
\frac 18\frac{L^{\prime\,2}/m^2}{1 + L^{\prime\,2}/m^2}
\right].
\end{equation}
At the limit $L^{\prime\,2} \gg m^2$
$$
M(p) \approx \frac{e^2}\pi m\left[\frac 34
\ln\left(\frac{L^{\prime\,2}}{m^2}\right) + \frac 38 \right],
$$
which coincides with a similar expression obtained in [4].

The components of impulse $L^\prime$ in (\ref{eq13}) form the Euclidean 4D
sphere $(L^{\prime\,2} = L_0^{\prime\,2} + \bm{L}^{\prime\,2})$
because of the rotation of the path of integration over variable $k^0$
in the complex plane. Besides, the components of impulse $L^\prime$ differ from
those of the previously introduced impulse $L$ because of the change in the
domain of integration from the replacement of variables, $k - p \to k$
\cite{ref4}.

To determine $L^\prime$, once again use equality
\[
\left\langle i \right|M(p)\left| i \right\rangle =
\left\langle i \right|Y_0(q)\left| i \right\rangle .
\]
Then,
\begin{equation}
\label{eq14}
\frac 34\ln\left(1 + \frac{L^{\prime\,2}}{m^2}\right) +
\frac{L^\prime}{2m}\arctan\frac m{L^\prime} -
\frac 18\frac{L^{\prime\,2}/m^2}
{1 + L^{\prime\,2}/m^2} = 1,115.
\end{equation}

The solution to equation (\ref{eq14}) is $L^\prime/m = 1.335 \approx 4/3,
\quad L^{\prime\,2} \approx 1.8m^2$.

With consideration of the above-described regularization of the expression
for the self-energy of free Dirac particle, the particle mass change due to
the renormalization in the second order of the perturbation theory is
\begin{equation}
\label{eq15}
\Delta m^{(2)} = \frac{e^2}\pi m \left(
\frac 32\ln(A_0)+\frac 1{4A_0^2}-\frac 14\right)=
1.115\frac{e^2}\pi m.
\end{equation}
The re-normalized particle mass is
\begin{equation}
\label{eq16}
m^{(2)} = m_0 + \Delta m^{(2)} = m_0\left(1 + 1.115\frac{e^2}\pi\right),
\end{equation}
where $m_0$ is the bare mass of the particle.

By setting the self energy of Dirac particle at rest,
$\left\langle i \right|Y_0(|\bm{p}| = 0)\left| i \right\rangle$,
equal to electrostatic energy, $e^2/r_{em}$, we obtain the
electromagnetic radius estimate for non-relativistic particle:
$r_{em} \approx 1/2m$.

Given external fields $A^\mu(\bm{x})$, the integration limit
$q$ and time component $L_0$ will depend on their magnitude with
retained invariant $L^2$.

For weak external fields $A^\mu( \bm{x})$, where in self-energy
expression (\ref{eq1}) we can restrict ourselves to inclusion only
of terms not higher than quadratic in the generalized particle impulse,
the results of the calculation for the anomalous magnetic moment of the
particle and Lamb shift of energy levels in the second order of the
perturbation theory within the approach developed in this paper will be
presented in the following publication.

In the presence of high external electromagnetic fields, electromagnetic
radius $r_{em}$ will reduce owing to changes of the integration limit
$q$, thus ensuring quantum electrodynamics applicability up to
ultra-short distances.

Of course, it would be desirable that $q(A^\mu(\bm{x}))$
be obtained from first principles in the future .

Summarize the results of the paper.

1. An algorithm for regularization of self-energy expressions for Dirac
particle satisfying the requirements of the relativistic and gauge
invariance is proposed.

2. The expression for the upper integration limit $q$ is a
relatively slowly varying function of particle impulse within the
second-order formulas of the ``old'' perturbation theory for free motion.

For a particle at rest, $q=m$; for $p/m$ in the range
from 0 to 1,
\[
q \approx m\left(1 + 0.00864\frac{p^4}{m^4} - 0.00217\frac{p^6}{m^6}\right);
\]
for the ultra-relativistic case $|\bm{p}| \gg m,\, q \approx 1.5m$.

It is shown for 4D perturbation theory upon introduction of limiting
4-impulse $L^2 = L_0^2 - \bm{L}^2$ that with a large magnitude of
the time component, $L_0/m \gg 1$, the spatial values of
$L_i$ are limited and are the same as the components of the introduced
limits of integration $q_i$: $\bm{L}^2 = \bm{q}^2$.

For the conventional integration techniques for Feynman integrals in the
second order of the perturbation theory the limiting integration impulse is
evaluated:
$L^{\prime\,2} = L_0^{\prime\,2} + \bm{L}^{\prime\,2} \approx
16m^2/9$.

3. In the second order of the perturbation theory, the re-normalized mass of
Dirac particle is
\[
m^{(2)} = m_0 + \Delta m^{(2)} = m_0
\left(1 + 1.115\frac{e^2}\pi\right),
\]
where $m_0$ is the bare mass of the particle.

4. The electromagnetic radius of particle at rest that can be found by
equating the self-energy and the electrostatic energy, $e^2/r_{em}$, is
approximately half the Compton wavelength,
$r_{em} \approx 1/2m$.

\end{document}